\documentclass[traditabstract,twocolumn]{aa}

\usepackage{epsfig,graphicx,natbib}
\usepackage{longtable}
\usepackage{amssymb}

\def\nH2{\hbox{$n_\mathrm{H_2}$}}
\def\kms{\hbox{km\,s$^{-1}$}}
\def\PKS1830{\hbox{PKS\,1830$-$211}}
\def\cm-2{\hbox{cm$^{-2}$}}

\begin{document}

\title{Detection of extragalactic CF$^+$ toward \PKS1830}
\subtitle{Chemical differentiation in the absorbing gas}

\author{S.~Muller \inst{1}
\and K.~Kawaguchi \inst{2}
\and J.\,H.~Black \inst{1}
\and T.~Amano \inst{3}
}

\institute{Department of Earth and Space Sciences, Chalmers University of Technology, Onsala Space Observatory, SE-43992 Onsala, Sweden
\and Department of Chemistry, Okayama University, Tsushima-naka 3-1-1, Kitaku, Okayama, 700-8530, Japan
\and Department of Chemistry and Department of Physics and Astronomy, University of Waterloo, Waterloo, ON N2L 3G1, Canada
}

\date {Received  / Accepted}

\titlerunning{CF$^+$ toward \PKS1830}
\authorrunning{}

\abstract{We report the first extragalactic detection of CF$^+$, the fluoromethylidynium ion, in the z=0.89 absorber toward \PKS1830. We estimate an abundance of $\sim$3$\times$10$^{-10}$ relative to H$_2$ and that $\sim$1\% of fluorine is captured in CF$^+$. The absorption line profile of CF$^+$ is found to be markedly different from that of other species observed within the same tuning, and is notably anti-correlated with CH$_3$OH. On the other hand, the CF$^+$ profile resembles that of [C\,I]. Our results are consistent with expected fluorine chemistry and point to chemical differentiation in the column of absorbing gas.}
\maketitle

\section{Introduction}

The CF$^+$ molecule was discovered in space by \cite{neu06} toward the Orion Bar region. It has been recently identified in Galactic diffuse clouds in absorption toward the background continuum sources 3C111, BL\,Lac, and W49 (\citealt{lis14,lis15}). It is also detected toward a high-mass protostar (\citealt{fec15}) and is therefore an ubiquitous species in the interstellar medium (ISM).

The chemistry of CF$^+$ and other fluorine-bearing species has been investigated by \cite{neu05} and \cite{neu09}. With an ionization potential higher than that of hydrogen, fluorine is mostly neutral in the diffuse ISM. Besides, it is a special element in having an exothermic reaction with H$_2$. With increasing H$_2$ fractional abundance, HF thus becomes the main interstellar reservoir of fluorine (F+H$_2$ $\rightarrow$ HF+H). HF is destroyed by reactions with He$^+$, H$_3^+$ and C$^+$. The latter reaction forms CF$^+$ (HF+C$^+$ $\rightarrow$ CF$^+$+H), which is destroyed by dissociative recombination. Eventually, CF$^+$ is expected to account for a small fraction ($\sim$1\%, \citealt{neu09}) of the gas-phase fluorine.

Recently, H$_2$Cl$^+$ was detected in the z=0.89 molecular absorber toward the lensed blazar \PKS1830 (\citealt{mul14b}), making it a good case to search for other halogen-bearing species. The absorber toward \PKS1830\ consists in two lines of sight, intercepting the disk of a z=0.89 face-on spiral galaxy (\citealt{wik96,win02}). Molecular absorption is seen in both lines of sight, with H$_2$ column density of $\sim$2$\times$10$^{22}$\,\cm-2\ toward the South-West image of the blazar, and $\sim$1$\times$10$^{21}$\,\cm-2\ toward the North-East image. More than 40 molecules have been detected in the SW line of sight (e.g., \citealt{mul11,mul14a}). The fractional abundances in the absorber are typical of those in-between diffuse and translucent Galactic clouds. Because the density is moderate, n(H$_2$)$\sim$ 10$^3$\,cm$^{-3}$ toward SW and likely even lower toward NE, the excitation of polar molecules is subthermal and radiatively coupled to the cosmic microwave background (CMB). Hence their rotation temperature is equal to 5.14\,K, the CMB temperature at z=0.89 (\citealt{mul13}). As a result the absorption spectrum toward \PKS1830\ is not too crowded (only low-energy levels are populated), and line identification is easy.

Here, we report the ALMA detection of CF$^+$ at z=0.89 toward \PKS1830, which is the first extragalactic detection of this species.

\section{Observations} \label{sec:data}

The observations of the CF$^+$ J=2-1 line, redshifted to 108.8\,GHz at z$_{abs}$=0.88582 (heliocentric frame), were taken during the ALMA Early Science Cycle~2 phase on 2014 August 27th. Two execution blocks of 6 minutes on source were observed. The correlator was configured with four different spectral windows of 1.875\,GHz, each counting 3840 channels separated by 0.488\,MHz, providing a velocity resolution of $\sim$2.6\,\kms\ after Hanning smoothing.

The data calibration was done within the CASA\footnote{http://casa.nrao.edu/} package, following a standard procedure. The bandpass response of the antennas was calibrated from observations of the bright quasar J\,1924$-$292. The quasar J\,1832$-$2039 was used for primary gain calibration. We further improved the data quality using self-calibration on the bright continuum of \PKS1830. The spectra were extracted toward both lensed images of \PKS1830, using the CASA-python task UVMULTIFIT (\citealt{mar14}) to fit a model of two point sources to the visibilities. We note that the two lensed images of \PKS1830, 1$''$ apart, were clearly spatially resolved, with a synthesized beam of 0.75$''$$\times$0.55$''$ (Briggs weighting with robust parameter set to 0.5) in the CLEAN-deconvolved maps.

\section{Results and discussion}

\begin{figure}[h] \begin{center}
\includegraphics[width=8.8cm]{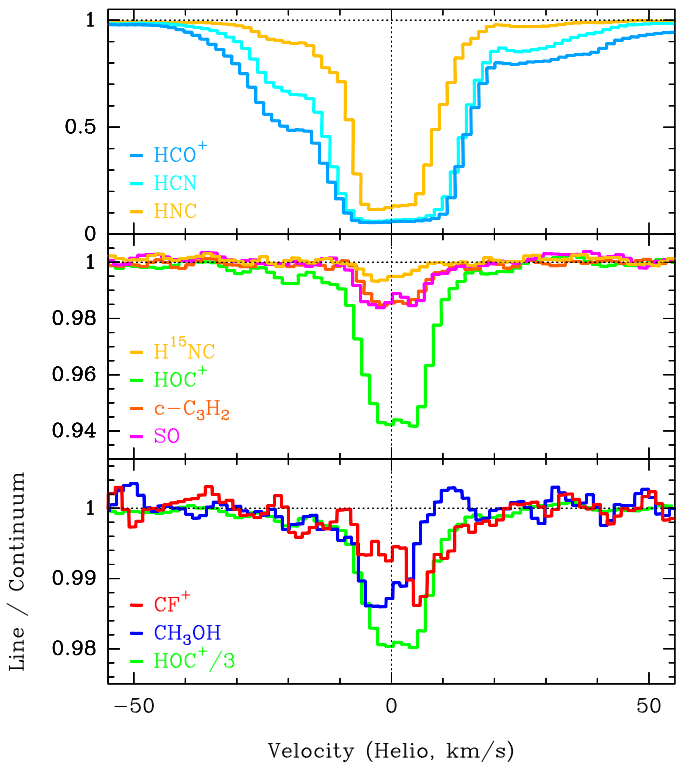}
\caption{Spectra of CF$^+$ and other species observed simultaneously toward the SW image of \PKS1830.}
\label{fig:spec-CF+}
\end{center} \end{figure}

The spectrum of CF$^+$ J=2-1 toward the SW image of \PKS1830\ is shown in Fig.\,\ref{fig:spec-CF+}, together with some other lines observed within the same tuning, namely from HCO$^+$, HCN, HNC, HOC$^+$, c-C$_3$H$_2$, SO, and CH$_3$OH. All these lines arise in low energy levels (see Table\,\ref{tab:line}). CF$^+$ is not detected toward the NE image (down to a rms of $\sim$0.001 of the continuum level per 2.6\,\kms\ channel), where the absorbing gas is found to be more atomic and diffuse than along the SW line of sight. This confirms that CF$^+$ is tracing gas with already substantial molecular fraction. We will restrict the discussion to the SW line of sight.

Despite having only one transition of CF$^+$, its identification is robust. As already said, the line crowding toward \PKS1830\ is low and we could not find any other possible alternative identification, except $^{13}$CN N=1-0 at z=0, but we would expect 1) the line to be very narrow (a few \kms, see the Galactic absorption of c-C$_3$H$_2$ toward \PKS1830\ by \citealt{sch15}) and decomposed into its characteristic spin-rotation and hyperfine pattern, and 2) very weak, as even the HCO$^+$/HCN/C$_2$H J=1-0 Galactic absorptions toward \PKS1830\ are no more than a few percent deep (Muller et al. in prep).

\begin{table*}[ht]
\caption{List of lines detected at z=0.89 toward the SW image of \PKS1830\ and discussed in this paper.}
\label{tab:line}
\begin{center} \begin{tabular}{lcccccccc}
\hline
          & Rest      & Sky freq.   & $E_{low}$ $^{(a)}$ & $S_{ul}$ $^{(b)}$ & Dipole   & $\alpha$ $^{(c)}$               & $\int \tau dv$ $^{(d)}$ & $N_{col}$ $^{(e)}$\\
 Line     & freq.      & $z$=0.88582 & (K)      &         & moment  & ($10^{12}$\,cm$^{-2}$\,km$^{-1}$\,s) & (\kms) & ($10^{12}$\,\cm-2) \\
          & (GHz)     & (GHz)       &          &         & (Debye)  &                         & & \\
\hline

HCN (2-1) & 177.26111 & 93.997 & 4.3 & 2.0 & 2.99 & 3.5 & -- & --\\
HCO$^+$ (2-1) & 178.37506 & 94.588 & 4.3 & 2.0 & 3.90 & 2.1 & -- & --\\
SO (5$_4$-4$_3$) & 178.60540 & 94.710 & 15.9 & 4.9 & 1.54 & 141.8 & 0.24 & 34.0 \\
HOC$^+$ (2-1)    & 178.97205 & 94.904 & 4.3 & 2.0 & 2.77 & 4.1 & 1.04 & 4.3 \\
HNC (2-1)     & 181.32476 & 96.152 & 4.4 & 2.0 & 3.05 & 3.4 & 41.9 & 142.5 \\
c-C$_3$H$_2$($para$) (4$_{1,3}$-3$_{2,2}$) & 183.62362 & 97.371 & 16.1 & 2.4 & 3.43 & 59.7 & 0.23 & 13.7 \\
CF$^+$ (2-1) & 205.17052 & 108.796 & 4.9 & 2.0 & 1.13 & 26.2 & 0.21 & 5.5 \\
CH$_3$OH (1$_1$\,A$^+$-2$_0$\,A$^+$) & 205.79127 & 109.126 & 7.0 & 0.5 & 1.44 & 255.1 & 0.18 & 45.9 \\

\hline
\end{tabular} 
\tablefoot{Frequencies and molecular data are taken from the Cologne Database for Molecular Spectroscopy (\citealt{mul01}).\\
$(a)$ $E_{low}$ is the lower-level energy of the transition. $(b)$ $S_{ul}$ is the line strength. $(c)$ $\alpha$ is defined as the conversion factor between integrated opacity and total column density: $N_{col}$=$\alpha \times \int \tau dv$, calculated under the assumption that the rotational excitation is coupled with the cosmic microwave background temperature, 5.14\,K at z=0.89 (see, e.g., \citealt{mul13}). $(d)$ Integrated opacity, over the velocity interval [$-$20;+20]\,\kms. The associated uncertainty is of 0.003 at 1$\sigma$. $(e)$ Total column density in the same velocity interval.}
\end{center} \end{table*}

The hyperfine structure in the rotational spectrum of CF$^+$ has been explored by \cite{guz12}. For the J=2-1 transition, the maximum spread between hyperfine components is less than 1\,\kms\ in the rest frame (this spread is to be divided by 1.89 due the redshift of the absorption), i.e., much narrower than the absorption line width ($\sim$15\,\kms, FWHM). We therefore neglect the hyperfine structure splitting in the following.

\begin{figure}[h] \begin{center}
\includegraphics[width=8.8cm]{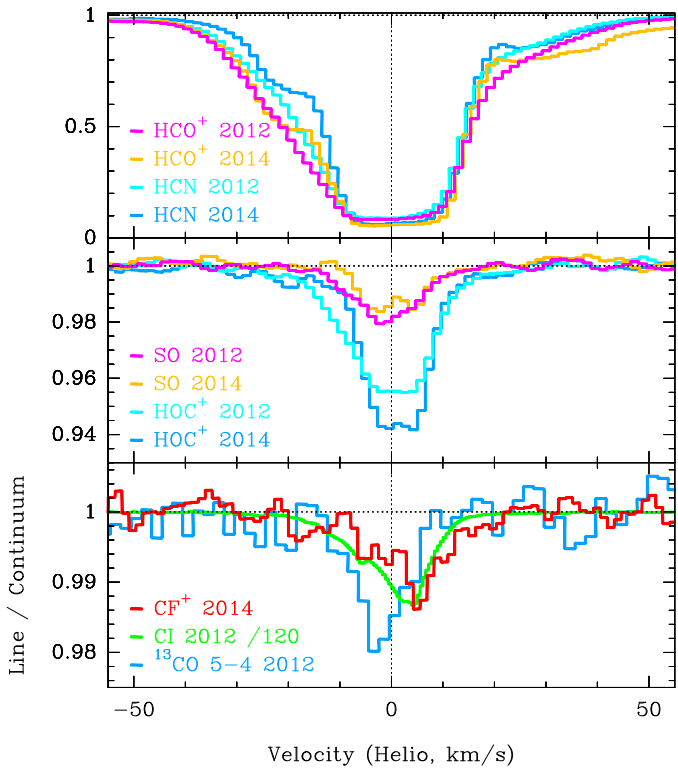}
\caption{Comparison of spectra obtained in 2014 and previous ALMA Cycle 0 spectra observed in 2012.}
\label{fig:spec-CF+timvar}
\end{center} \end{figure}

The continuum source covering factor $f_c$ can be inferred from the saturation level of the HCO$^+$ and HCN J=2-1 lines, yielding $f_c$$\sim$94\% \footnote{We note that this value is slightly higher than that observed in 2012 ($\sim$91\%) by \cite{mul14a}, see Fig.\,\ref{fig:spec-CF+timvar}.}. This is a strong indication that weak absorption of a few percent of the continuum level corresponds to optically thin lines. The CF$^+$ absorption, at most $\sim$1.5\% of the continuum level, is thus likely to be optically thin. However, it does not resemble that of any other species observed within the same tuning (Fig.\,\ref{fig:spec-CF+}). In fact, we can classify the species in four categories depending on their line profiles: 1) HCO$^+$, HCN, and HNC, with their optically thick absorption (HCO$^+$ and HCN reach saturation); 2) HOC$^+$, c-C$_3$H$_2$, and SO, which are optically thin and trace the bulk of the main absorption between about $-$20\,\kms\ \footnote{Throughout the paper,we refer the velocity to the heliocentric frame, taking z$_{abs}$=0.88582.} to $+$20\,\kms; 3) CF$^+$, which peaks on the red wing at about $+$5\,\kms, although its absorption is also seen over the main $-$20\,\kms\ to $+$20\,\kms\ velocity range; and 4) CH$_3$OH which peaks on the blue wing at about $-$3\,\kms.

It is interesting to note that in previous observations toward \PKS1830, only few species show slight but significant deviations in their velocity centroids compared to the majority of detected species. In the 7\,mm spectral scan obtained with the Australia Telescope Compact Array (ATCA), \cite{mul11} found that HCO and SO$^+$ had a redshifted centroid ($\sim$+4\,\kms) \footnote{Note the limited velocity resolution of the ATCA 7\,mm survey, of $\sim$10--6\,\kms\ across the 30--50\,GHz frequency interval.}, similar to CF$^+$ here. On the other hand, CH$_3$OH, HNCO, HC$_3$N, and CH$_3$NH$_2$ were found with blueshifted centroids ($\sim$$-$5\,\kms), like CH$_3$OH here. A similar centroid was found by \cite{kan15} on the line profile of several CH$_3$OH lines observed at good signal-to-noise ratio with the Jansky Very Large Array. The time variations of the absorption (\citealt{mul08,mul14a}) hamper the comparison of line profiles across time, but at a given epoch, the relative offsets of the line velocity centroids between different species are meaningful. Since all those lines have similar excitation, the offsets ought to originate from differences in the chemical or physical properties of the absorbing medium. In Fig.\,\ref{fig:spec-CF+timvar}, we compare the new spectra obtained in 2014 with those of the same lines observed in 2012 (\citealt{mul14a}, ALMA Cycle~0 data). For HOC$^+$, the 2014 spectrum is about 30\% deeper around $v$=0\,\kms\ and the blue wing has become slightly less prominent (same as for the HCO$^+$ and HCN profiles), but overall the line shape hasn't changed drastically. Most interestingly, of all species observed in 2012 by ALMA, it is [C\,I] which has the closest line profile to CF$^+$. Neutral carbon is naturally expected to be present in regions better shielded and slightly denser than for ionized carbon. Since CF$^+$ is produced from C$^+$, we expect an even better correlation between CF$^+$ and [C\,II] lines.

By integrating over the line between $-$20\,kms\ to +20\,\kms\ and assuming thermal equilibrium with CMB photons, we find a total CF$^+$ column density of 5.5$\times$10$^{12}$\,\cm-2. With a H$_2$ column density of 2$\times$10$^{22}$\,\cm-2\ (\citealt{mul14a}), this corresponds to an averaged fractional abundance [CF$^+$]/[H$_2$] of $\sim$3$\times$10$^{-10}$. The cosmic abundance of fluorine is a few times 10$^{-8}$ that of hydrogen (\citealt{sno07, asp09}). Assuming that the fluorine abundance is similar in the absorbing clouds of the z=0.89 galaxy, we conclude that only $\sim$1\% of the interstellar fluorine is captured in CF$^+$, in agreement with predictions from \cite{neu09}. The CF$^+$ production and destruction may be expressed as follows, in diffuse clouds, 
\begin{equation}
{\rm C}^+ + {\rm HF} \rightarrow {\rm CF}^+ + {\rm H},    
\end{equation}
with a formation rate $k_f$=7.2$\times$10$^{-9}$$\times$(T/300)$^{-0.15}$\,cm$^3$\,s$^{-1}$, T being the temperature, and
\begin{equation}
{\rm CF}^+ + {\rm e}  \rightarrow {\rm C} + {\rm F},     
\end{equation}
\noindent with a dissociative recombination rate $k_e$=5.2$\times$10$^{-8}$$\times$(T/300)$^{-0.8}$\,cm$^3$\,s$^{-1}$. There, other destruction reactions with negative ions are neglected. By assuming steady state for the CF$^+$ abundance, the following relation is obtained: 
\begin{equation} \label{eq:relativeCF+}
\frac{[{\rm CF}^+]}{[{\rm HF}]} = \frac{ k_f [{\rm C}^+]}{k_e[{\rm e}]}  \approx \frac{k_f}{k_e} = 0.138 \times \left ( \frac{T}{300} \right )^{0.65}
\end{equation}
\noindent where the electron density [e] is assumed to be roughly equal to the C$^+$ abundance in diffuse cloud. The recombination rate constant $k_e$ has been measured by using a laboratory storage ring experiment (\citealt{nov05}). On the other hand the $k_f$ value has not yet been measured. \cite{neu05} estimated the value by use of a statistical adiabatic capture model. Therefore, the coefficient 0.138 may have an uncertainty of a factor of 2--3. When we assume a kinetic temperature T=100 K, Eq.\,\ref{eq:relativeCF+} gives the ratio [CF+]/[HF]$\sim$7\%, where it is noted that the T value corresponds to roughly A$_{\rm V}$=2 from Fig.\,10 of the paper of \cite{nag13}. In more sophisticated calculations, the ratio changes depending on the A$_{\rm V}$ value, taking the maximum ratio of $\sim$10\% at A$_{\rm V}$=1, as shown in Fig.\,3 of \cite{neu09}. From the observational point of view, the CF$^+$ abundance is thought to be about 1\% of the gas-phase fluorine abundance in the Galactic diffuse clouds (e.g., \citealt{nag13}).

CF$^+$ and HF have both been detected toward the Galactic source W49 (\citealt{lis15}), and the abundance ratio [CF$^+$]/[HF] is in the range 0.010--0.025 for various velocity components. The values are consistent if we consider the uncertainty of the calculated $k_f$ (see also the discussion by \citealt{lis15}) and variations of C$^+$ abundance in diffuse clouds.

For the absorber toward \PKS1830, the HF J=1-0 line, redshifted into the ALMA band 9, has recently been detected (\citealt{kaw16}) and shows a heavily saturated absorption toward the SW image, with a HF column density $>$3.4$\times$10$^{14}$\,\cm-2.
The fluoronium ion, H$_2$F$^+$, on the other hand, was not detected down to an abundance ratio [H$_2$F$^+$]/[HF]$\leq$1/386 (3$\sigma$). This confirms that HF is the main molecular reservoir of fluorine.

\section{Summary and conclusions} \label{sec:conclusions}

We report the detection of CF$^+$ in the z=0.89 absorber toward \PKS1830. This is the first extragalactic detection of this ion which, by now, has been observed in various Galactic sources. We find that the absorption line profile of CF$^+$ is significantly different from those of other species observed in the same tuning, and especially anti-correlated with CH$_3$OH. We estimate a CF$^+$ fractional abundance of $\sim$3$\times$10$^{-10}$, implying that only $\sim$1\% of the interstellar fluorine is captured in this ion, and that HF is by far the main reservoir of interstellar fluorine in the molecular phase.

\begin{acknowledgement}
This paper makes use of the following ALMA data: ADS/JAO.ALMA\#2013.1.01099.S and ADS/JAO.ALMA\#2011.0.00405.S. ALMA is a partnership of ESO (representing its member states), NSF (USA) and NINS (Japan), together with NRC (Canada) and NSC and ASIAA (Taiwan) and KASI (Republic of Korea), in cooperation with the Republic of Chile. The Joint ALMA Observatory is operated by ESO, AUI/NRAO and NAOJ.
\end{acknowledgement}

\appendix

\section{Data}

\begin{figure*}[h] \begin{center}
\includegraphics[width=\textwidth]{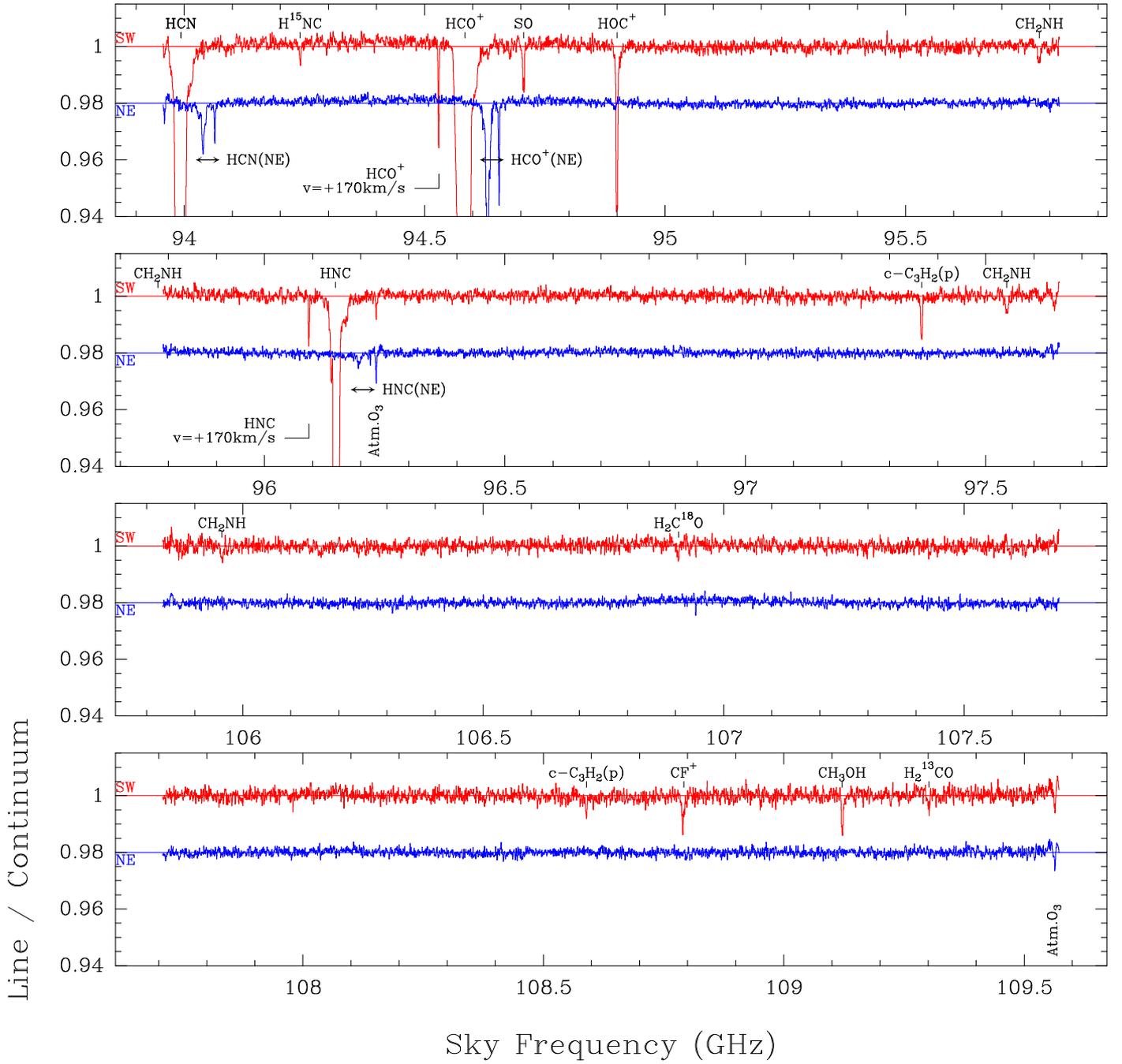}
\caption{Full ALMA spectra toward the SW (in red) and NE (in blue, offset by $-$0.02) images of \PKS1830. The spectra are normalized to the continuum level.}
\label{fig:full-spec}
\end{center} \end{figure*}

\end{document}